\documentclass[reprint,amsmath,amssymb,aps,longbibliography]{revtex4-2}

\pdfoutput=1
\usepackage{graphicx}
\graphicspath{{images/}}
\usepackage{float}
\usepackage{adjustbox}
\usepackage{verbatim}

\usepackage{upgreek}\usepackage{graphicx}% Include figure files
\usepackage{dcolumn}% Align table columns on decimal point
\usepackage{bm}% bold math
\usepackage{color}
\usepackage{mathtools}
\usepackage[normalem]{ulem}
\usepackage[english]{babel}
\usepackage{xr-hyper}
\usepackage[colorlinks=true,linkcolor=blue,urlcolor=blue,citecolor=blue]{hyperref}

\usepackage{amssymb}
\usepackage{mathrsfs}
\usepackage{amsmath,amsfonts,bm,graphicx}
\usepackage{upgreek}
\usepackage{color}
\usepackage{etoolbox}

\newcommand{\app}[1]{Appendix~\ref{#1}}
\newcommand{\sect}[1]{Sec.~\ref{#1}}
\newcommand{\fig}[1]{Fig.~\ref{#1}}
\newcommand{\Fig}[1]{Figure~\ref{#1}}
\newcommand{\eq}[1]{Eq.~(\ref{#1})}
\newcommand{\eqs}[2]{Eqs.~(\ref{#1}) and (\ref{#2})}
\newcommand{\eqr}[2]{Eqs.~(\ref{#1})-(\ref{#2})}
\newcommand{\Eq}[1]{Equation~(\ref{#1})}

\newcommand{\hd}[1]{{\bf #1:~}}

\newcommand{\roundbk}[1]{\left({#1}\right)}
\newcommand{\squarebk}[1]{\left[{#1}\right]}

\newcommand{\N}{\mathcal{N}}

\newcommand{\Vkl}{V}
\newcommand{\PIS}{P_{\text{IS}}}
\newcommand{\FIS}{F_{\text{IS}}}

\newcommand{\rev}[1]{{#1}}

\begin{document}
\title{Distinguishable-particle Glassy Crystal: the simplest molecular model of glass}
\author{Leo S.I. Lam$^1$}
\thanks{These authors contributed equally}
\author{Gautham Gopinath$^{2,3}$}
\thanks{These authors contributed equally}
\author{Zichen Zhao$^{2,4}$}
\thanks{These authors contributed equally}
\author{Shuling Wang$^{5,2}$}
\author{Chun-Shing Lee$^{6,2}$}
\author{Hai-Yao Deng$^7$}
\author{Feng Wang$^8$}
\author{Yilong Han$^8$}
\author{Cho-Tung Yip$^6$}
\email[Email: ]{h0260416@gmail.com}
\author{Chi-Hang Lam$^2$}
\email[Email: ]{C.H.Lam@polyu.edu.hk}
\address{
  $^1$Department of Computer Science, University of Warwick, Coventry, UK \\
  $^2$Department of Applied Physics, Hong Kong Polytechnic University, Hong Kong, China \\
  $^3$Department of Physics, Yale University, New Haven, Connecticut 06511, USA. \\
  $^4$Department of Physics, North Carolina State University, Raleigh, NC 27695, USA \\
  $^5$School of Mathematics and Physics, Hebei University of Engineering, Handan City 056038, China\\
  $^6$Department of Physics, Harbin Institute of Technology, Shenzhen 518055, China\\
  $^7$School of Physics and Astronomy, Cardiff University, 5 The Parade, Cardiff CF24 3AA, Wales, UK\\
  $^8$Department of Physics, Hong Kong University of Science and Technology, Clear Water Bay, Hong Kong, China
}
\begin{abstract}
The nature of glassy dynamics and the glass transition are long-standing problems  under active debate. In the presence of a structural disorder widely believed to be an essential characteristic of structural glass, identifying and understanding  key dynamical behaviors are very challenging. In this work, we demonstrate that an energetic disorder, which usually results from a structural disorder, is instead a more essential feature of glass. Specifically, we develop a distinguishable-particle glassy crystal (DPGC) in which particles are ordered in a face-centered cubic lattice and follow particle-dependent random interactions, leading to an energetic disorder in the particle configuration space. Molecular dynamics simulations in the presence of vacancy-induced particle diffusion show typical glassy behaviors. A unique feature of this molecular model is the knowledge of the complete set of inherent structures with easily calculable free energies, implying a well-understood potential energy landscape. 
\end{abstract}

\maketitle

\label{sec:introduction}
There are many open questions regarding the dynamics of structural glass and the nature of the glass transition despite decades                                                     
of intensive study \cite{stillinger2013review,biroli2013review,mckenna2017,arceri2022}.
Glass formers constitute an immensely diverse group of materials. Besides many fascinating phenomena in the bulk, they also exhibit puzzling features in confined geometries \cite{tsui2014review,roth2021review}. A typical glass is characterized by a structural disorder with random and frustrated particle positions and, for non-spherical molecules, also orientations.
A type of glass, called orientational glassy crystal, possesses a crystalline structure but has random molecular orientations which constitute a structural disorder \cite{hochli1990review}.
In view of the wide range of materials and physical conditions, identifying and understanding the most fundamental features of glass have proven very challenging.

In general, the structural disorder of glass amounts to momentarily quenched random particle separations and (or) orientations. This implies random particle interacations and hence also an energetic disorder. An important question is whether structural disorder plays other crucial roles in glassy dynamics apart from generating the energetic disorder. By studying a distinguishable-particle glassy crystal (DPGC), we show that structural disorder plays no other role in many glassy properties. The DPGC is a molecular model of glass with an energetic disorder, but it is structurally ordered with neither positional nor orientational disorder, except for isolated point defects. The DPGC is in our knowledge the simplest molecular model of glass. It is characterized by vacancy-induced dynamics. The inherent structures are trivially known. it is also the only known molecular model with solvable equilibrium statistics. It can be an ideal example for studying glassy dynamics.

The model is a direct molecular generalization of the distinguishable-particle lattice model (DPLM), which has successfully reproduced many glassy phenomena \cite{zhang2017,lee2021,lulli2020,lee2020,gopinath2022,  lulli2021,gao2022,gao2023,ong2024,zhai2024} including non-trivial ones such as Kovacs paradox \cite{lulli2020}, a wide range of fragilities \cite{lee2020}, and diffusion coefficient power-laws upon partial swap \cite{gopinath2022}. The energetics in the DPLM is dictated by particle-dependent nearest neighbor pair interactions and the kinetics is characterized by void-induced particle hops \cite{lee2020}. It requires no  explicit kinetic or energetic constraint \cite{garrahan2011review, ritort2003review, biroli2001}, enabling a generalization to a molecular model.

We consider dynamics dominated by vacancy-induced particle hops. Voids, the counterpart of vacancy in glass, or related defects have long been suggested to be responsible for glassy dynamics \cite{glarum1960,fredrickson1986,napolitano2013,lam2017}.
Recently, quasivoids, i.e. voids in a fragmented form, have been identified to dominate dynamics in glassy colloidal experiments \cite{yip2020} and locally averaged free volume has been found to correlate with dynamics in hard spheres simulations \cite{mei2022}.   These motivate the use of vacancy-induced dynamics in this work, although the importance of quasivoids to glass in general is still an open question.

The rest of the paper is organized as follows. In \sect{sec:model} we will explain in detail the DPGC model. In \sect{sec:results} and \sect{sec:dyn} we demonstrate standard glassy behaviors and the dynamic heterogeneity of DPGC respectively. Then we explain the inherent structures and their equilibrium statistics in \sect{sec:eq} and conclude in \sect{sec:Conclusion} with further discussions.

\section{Model}
\label{sec:model}

\begin{figure}[tb]
\includegraphics[width=\linewidth]{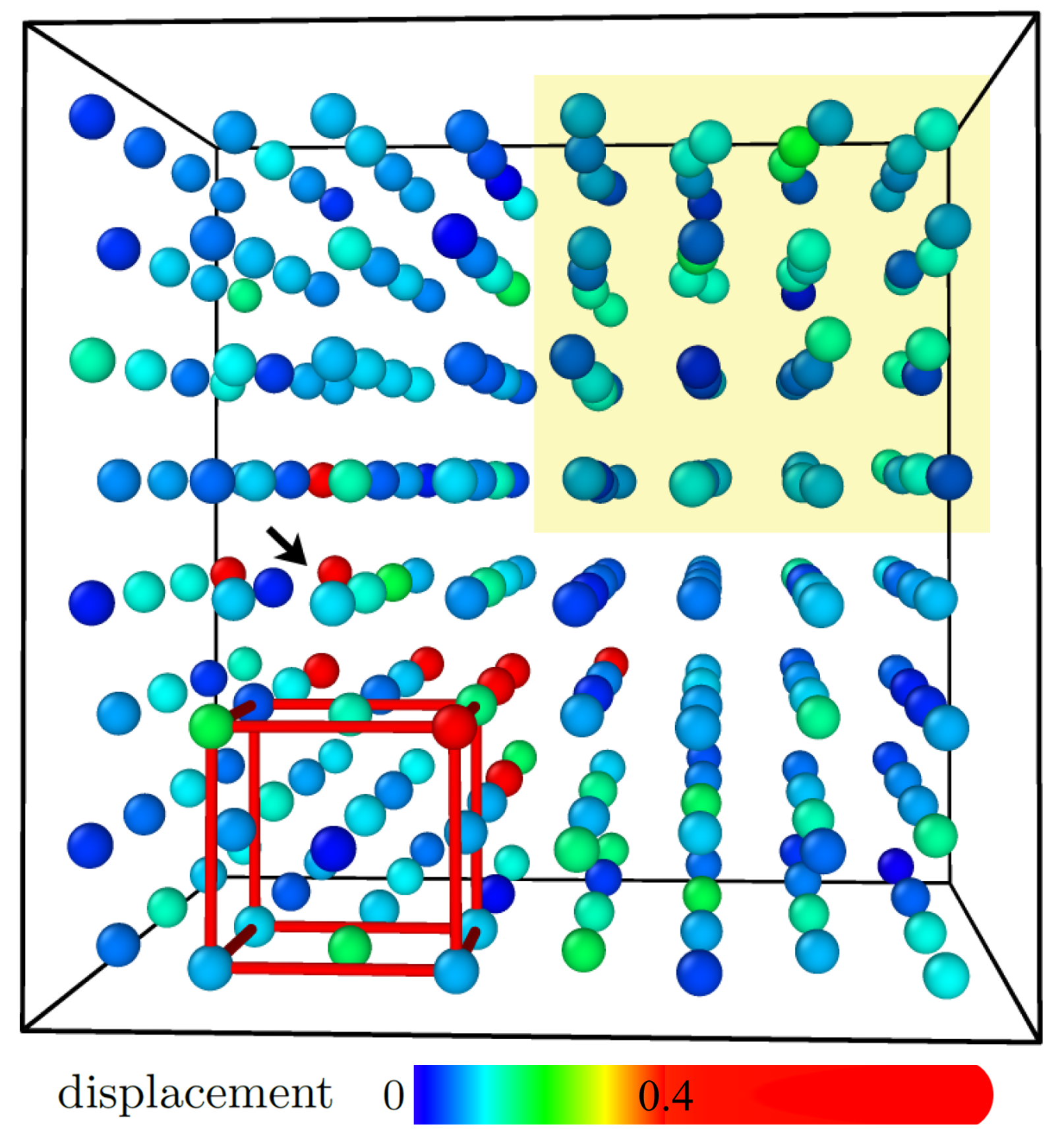}
\caption{
  A snapshot of a small-scale DPGC with 255 particles and one vacancy (black arrow) following an FCC structure in a cubic simulation box at $T=0.3$. Particle colors indicate displacements from 0 to 0.4$\sigma$  over
a duration of $2 \times 10^5$. Particles with displacements beyond 0.4$\sigma$ (red) have hopped at least once.
The majority of particle movement is found near the vacancy.
Particles at the top right corner (yellow region) are shown at their instantaneous positions. Other particles are shown at their inherent structural positions. 
A small red cube represents the FCC unit cell. }
\label{fig:lattice}
\end{figure}

In the DPGC, $N$ distinguishable particles form a  face-centered cubic (FCC) lattice in 3D (see yellow region in \fig{fig:lattice}). The interaction energy $\Phi_{kl}(r)$ between particles $k$ and $l$ separated by a distance $r$ follows the Lennard-Jones (LJ) potential
\begin{equation}
\Phi_{kl}\left(r\right)=-4V_{kl}\left[\left(\frac{\sigma}{r}\right)^{12}-\left(\frac{\sigma}{r}\right)^{6}\right]
\end{equation}
with a distance cut off of 2.5$\sigma$ beyond which it becomes a constant with respect to $r$. We fix $\sigma = 1$ which defines the length scale in our system. Each $V_{kl}<0$ represents a particle-dependent energy depth of the LJ potential and is a quenched random variable following a uniform distribution $g(V)$ in the range [-1, -0.25] \cite{rabin2015}. The random depth $V_{kl}$ is analogous to random interactions in the DPLM which follows a range $[-0.5, 0.5]$ designed to minimize effective vacancy-vacancy attraction \cite{lee2020}. %To ensure a hard-core repulsion, this range is chosen as it provides sufficiently negative energy depths, which however also induce a considerable vacancy-vacancy attraction. A small vacancy density is thus adopted to avoid vacancy aggregation.
%  This range is chosen such that it provides sufficiently negative energy depths to ensure hard-core repulsion. This however induces a considerable vacancy-vacancy attraction \cite{zhang2017}.
This range is not appropriate for the DPGC as $V_{kl}$ must now be negative. To avoid vacancy aggregation, we instead have to employ a small vacancy density.

Our main molecular dynamics (MD) simulations are performed in the NVT ensemble in a cubic box with $13^3$ FCC unit cells under periodic boundary conditions. The lattice points are  occupied by $N=8780$ particles with $N_v=8$ vacancies, corresponding to a vacancy density of $\phi_v \simeq 0.091\%$ per lattice point.  
The lattice constant is set at $a_0=1.6\sigma$. This  implies a nearest-neighbor distance of $a_0/\sqrt{2} \simeq 1.131\sigma$ and a small tensile strain which has been found necessary to break a form of divacancy-particle complex. The FCC structure is remarkably stable below the melting temperature of about $0.83$ (see \app{melting}). In contrast, a disordered form of the system is stable only if a bimodal distribution of particle diameters is adopted \cite{qin2023}.
The DPGC can be brought to equilibrium using swap and a ghost particle method (see \app{fastequil}).

\section{Glassy characteristics}
\label{sec:results}

We now explain main measurements on our MD simulations of the DPGC demonstrating glassy behaviors, while more details will be elaborated in subsequent sections.

\begin{figure}[tb]
  \includegraphics[width=0.96\columnwidth]{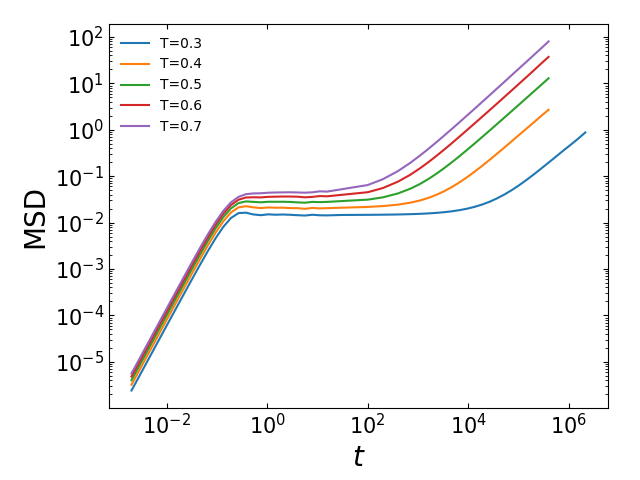}
  \vspace{-0.07\columnwidth}
  \caption{MSD in log-log plot against time.}
  \label{fig:MSD}
\end{figure}
  
 \begin{figure}[tb]
  \includegraphics[width=0.96\columnwidth]{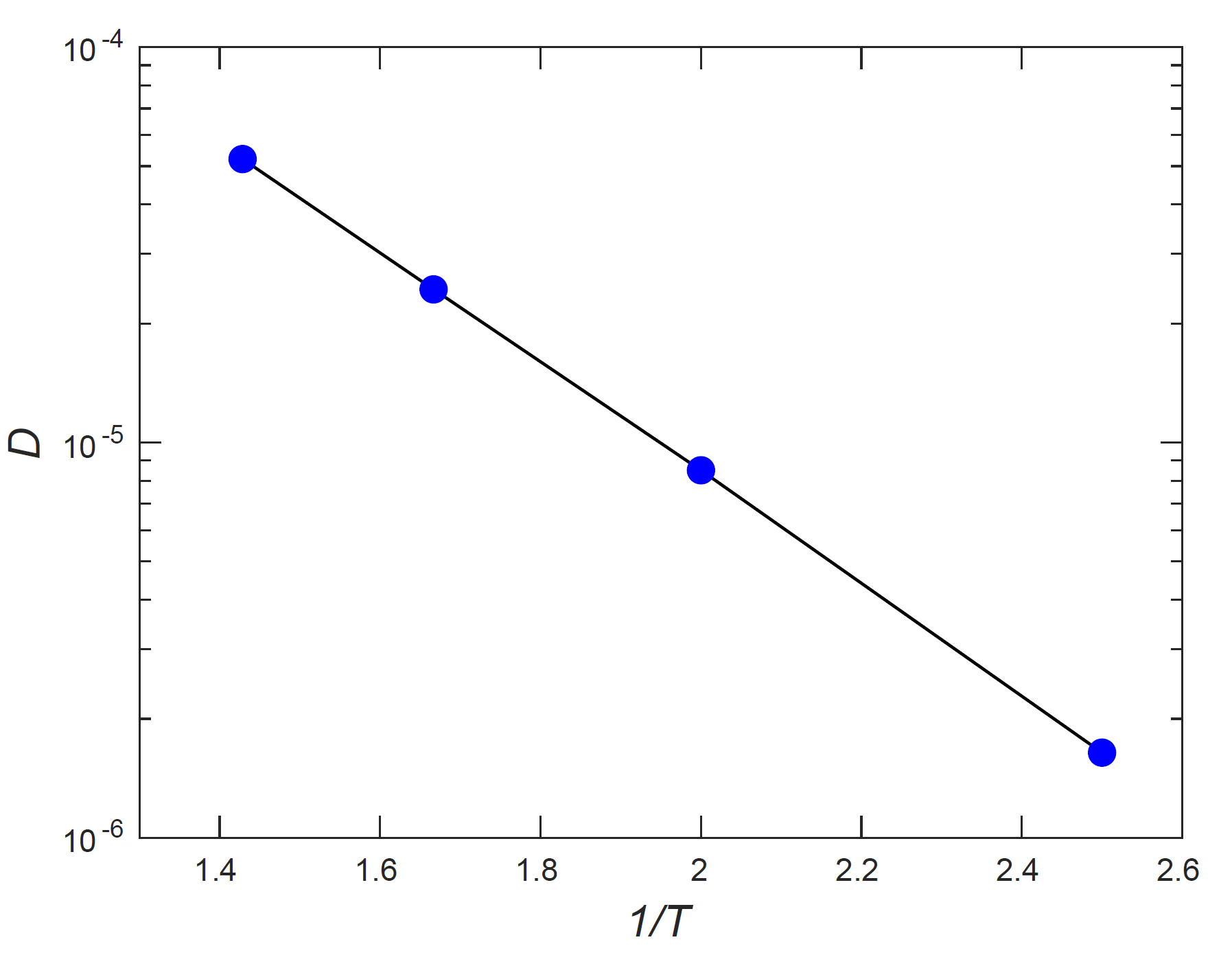}
  \vspace{-0.04\columnwidth}
  \caption{Semi-log plot of diffusion coefficient $D$, showing the Arrhenius relation \rev{for $T=0.4$ to $T=0.7$.}}
  \label{fig:d}
\end{figure}
  
\hd{Mean-squared displacement}
We calculate the particle mean-squared displacement (MSD) defined as $\left\langle |\mathbf{r}_l(t) - \mathbf{r}_l(0)|^2 \right\rangle$
where $\mathbf{r}_l(t)$ denotes the position of particle $l$ at time $t$ \cite{kawasaki2013}.
\Fig{fig:MSD} shows the MSD in a log-log plot for different temperature $T$. The plateau before the diffusive regime shows typical glassy characteristics.
At long time in the diffusive regime when the MSD is beyond $\sigma$,
we measure the particle diffusion coefficient from
\begin{equation}
  D = \frac{1}{2d}  \frac{MSD}{t}
\end{equation}
where $d=3$ indicates three-dimensional space. Results are shown in \fig{fig:d}.

\begin{figure}[tb]
  \includegraphics[width=\linewidth]{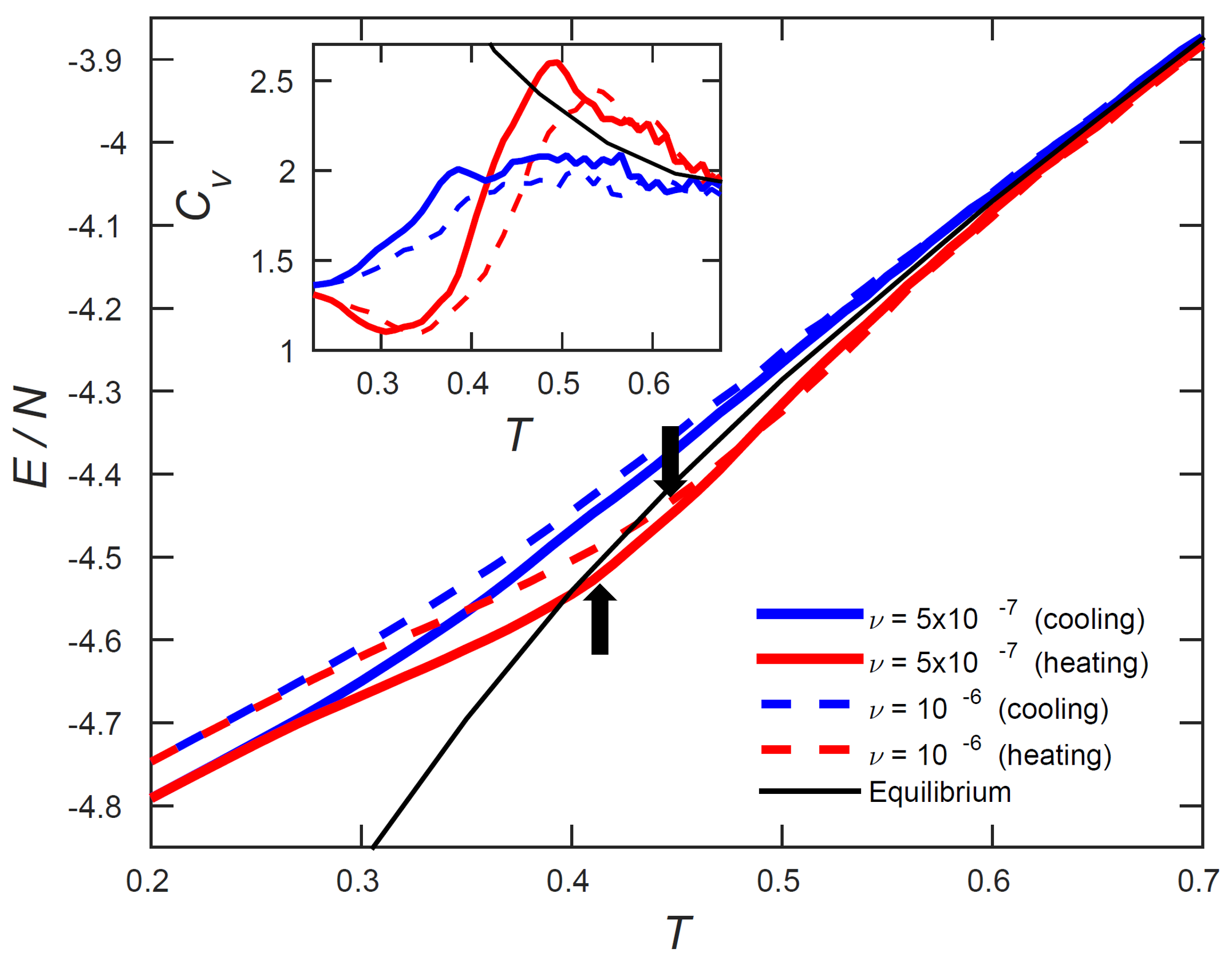}
  \caption{
  Potential energy per particle $E/N$ against temperature $T$ during cooling (blue) and heating (red) with rates $\nu=5\times10^{-7}$ and $10^{-6}$.  The black line shows the equilibrium energy. Black arrows show $T_g$ for both rates. Inset shows heat capacity $C_v$ against $T$.
  }
  \label{fig:cool}
\end{figure}

\hd{Energy Hysteresis} 
An energy hysteresis is observed during a cooling/heating cycle. Starting from an equilibrium system at temperature $0.7$, it is cooled to $0.2$ and then heated back to $0.7$ at the same cooling/heating rate $\nu$.
\Fig{fig:cool} plots the average potential energy per particle $E/N$ against temperature $T$ for two values of $\nu$. 
We observe a clear energy hysteresis with kinks signifying glass transitions. 

\hd{Glass transition temperature}
The glass transition temperature $T_g$ can be found from the intersection of the two relatively linear sections of the heating curves \cite{hodge1994}. Using data from \fig{fig:cool}, we get $T_g\simeq0.41$ and 0.44 for $\nu=5\times10^{-7}$ and $10^{-6}$ respectively, showing that $T_g$ increases with $\nu$ as expected of glass.

The glass transition temperature $T_g$ separates two phases of the system exhibiting  different dynamics. Below $T_g$, the system enters the glass phase with largely frozen particle allocations to the lattice points. Particle motions are mainly vibrations and back-and-forth hops. Above $T_g$, particles perform vacancy-induced hops around various lattice positions within practical observation times. This rearranges the particles in the lattice and relaxes the pair interactions. We propose to call it the dynamic phase, which is analogous to the liquid phase of conventional glass formers.  Importantly, such energy relaxation in the dynamic phase, akin to glassy materials, is absent in conventional monoatomic crystals.

\begin{figure}[tb]
  \includegraphics[width=\linewidth]{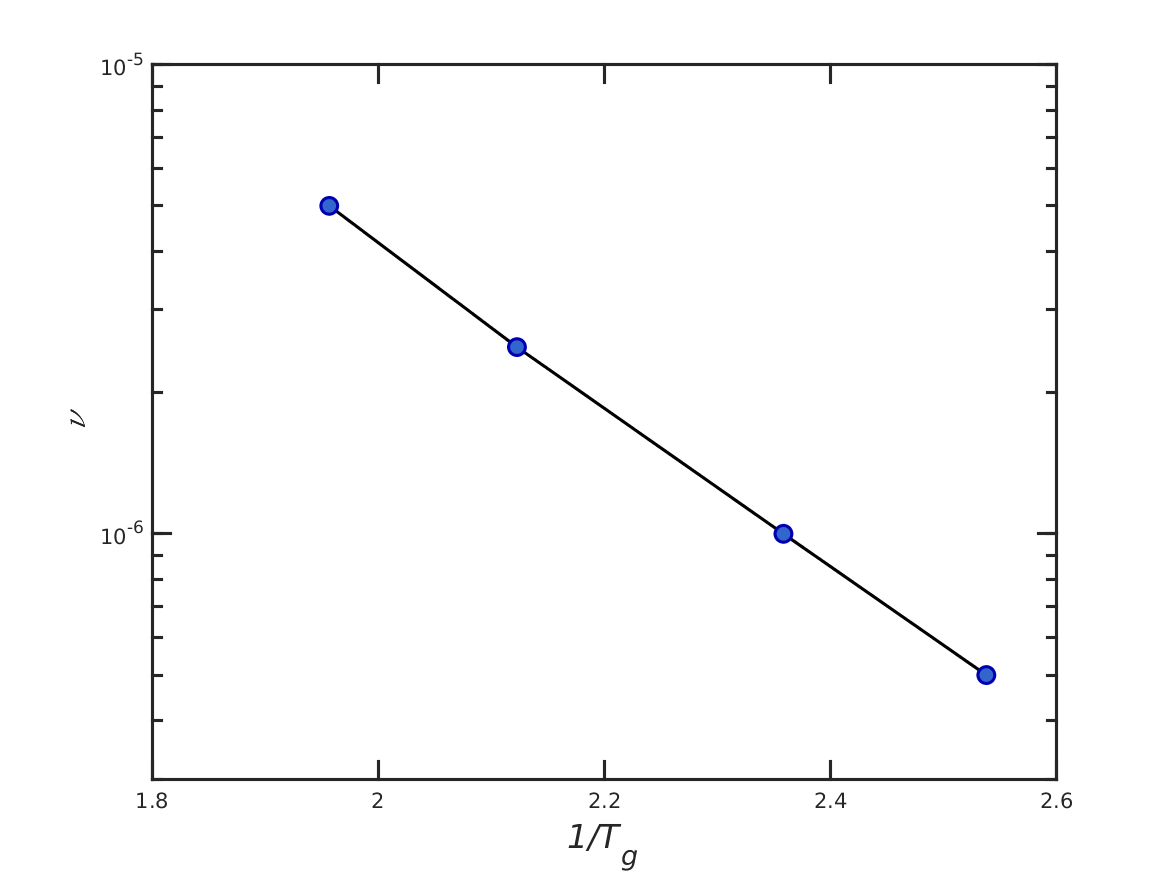}
\caption{A semi-log plot of cooling rate $\nu$ against $1/{T_{g}}$ based on MSD measurements.}
\label{fig:Tg}
\end{figure}

Recognizing that in the glass phase, particles hop infrequently, we measure $T_g$ based on particle MSD, which provides better statistics. When cooling the system from $0.7$  to a low temperature $T_0=0.2$ at a rate $\nu$, we measure the MSD from the particle position ${\bf r}(T)$ at temperature $T$ to the final frozen position ${\bf r}(T_0)$, i.e. MSD =  $\langle |{\bf r}(T)-{\bf r}(T_{0})| ^2\rangle$. We define $T_{g}$ as the temperature at which on average, all particles have hopped away from their initial positions once, to a neighboring lattice point before becoming frozen, i.e. MSD = $ a_{0}^2/2$.
\Fig{fig:Tg} plots $\nu$ against the measured $1/T_g$. The linearity in the semi-log plot shows  
$\log(\nu) \sim 1/T_g$, consistent with typical glassy characteristics \cite{hodge1994}.
In particular, it indicates that the present DPGC is a strong glass. The glassy crystal studied shows relaxation behaviors of strong glass due to the choice of an uniform interaction energy distribution $g(V)$. A more fragile relaxation is observed under a bi-component form of interaction energy distribution, in agreement with results on the DPLM \cite{lee2020}. This is explained in detail in \app{fragile}.

\begin{figure}[tb]
  \flushleft{~(a)}\\ \vspace{-0.1\columnwidth} ~~~~~~
  \includegraphics[width=\linewidth]{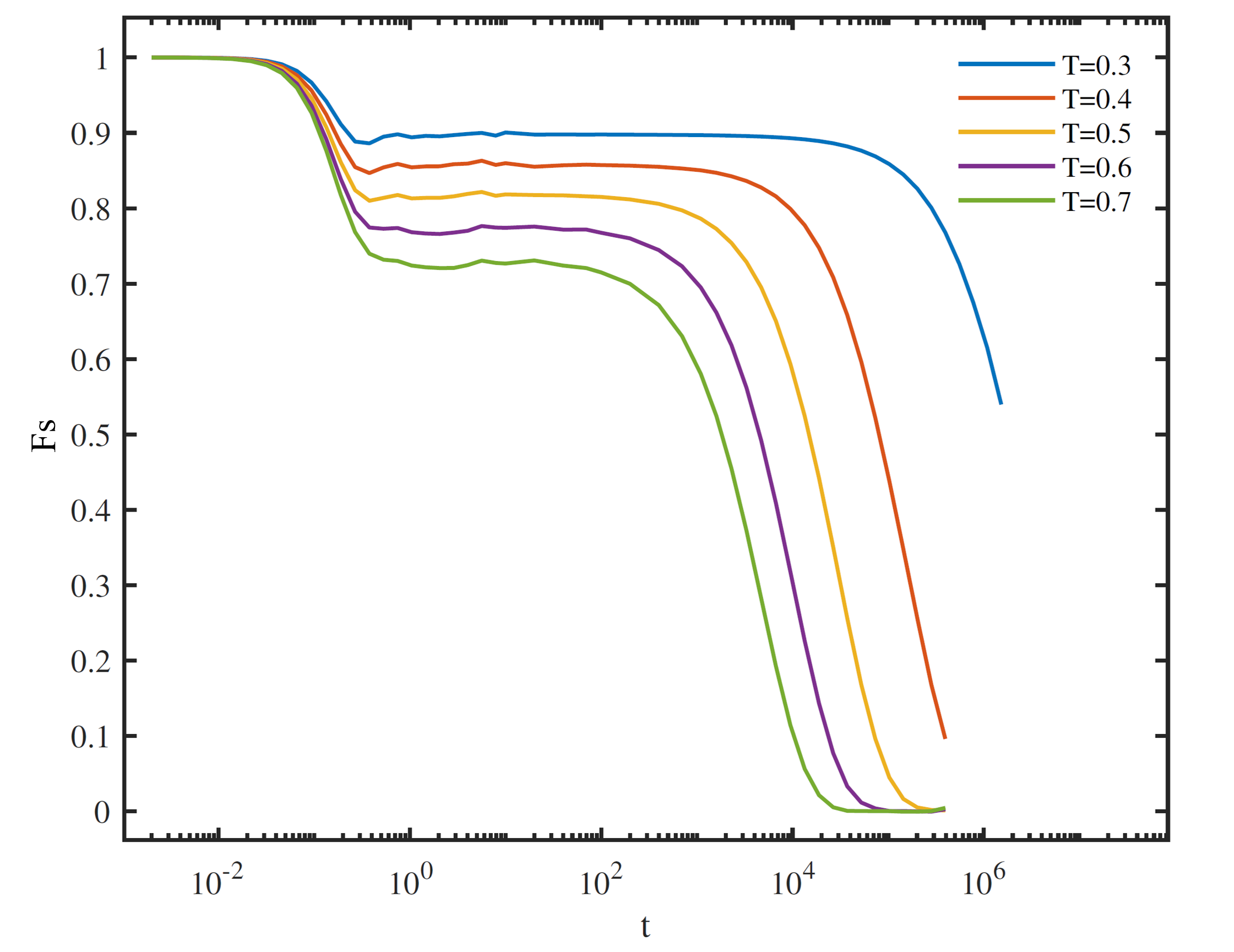}
  \vspace{-0.04\columnwidth}
  \flushleft{~(b)}\\ \vspace{-0.1\columnwidth} ~~~~~~
  \includegraphics[width=\linewidth]{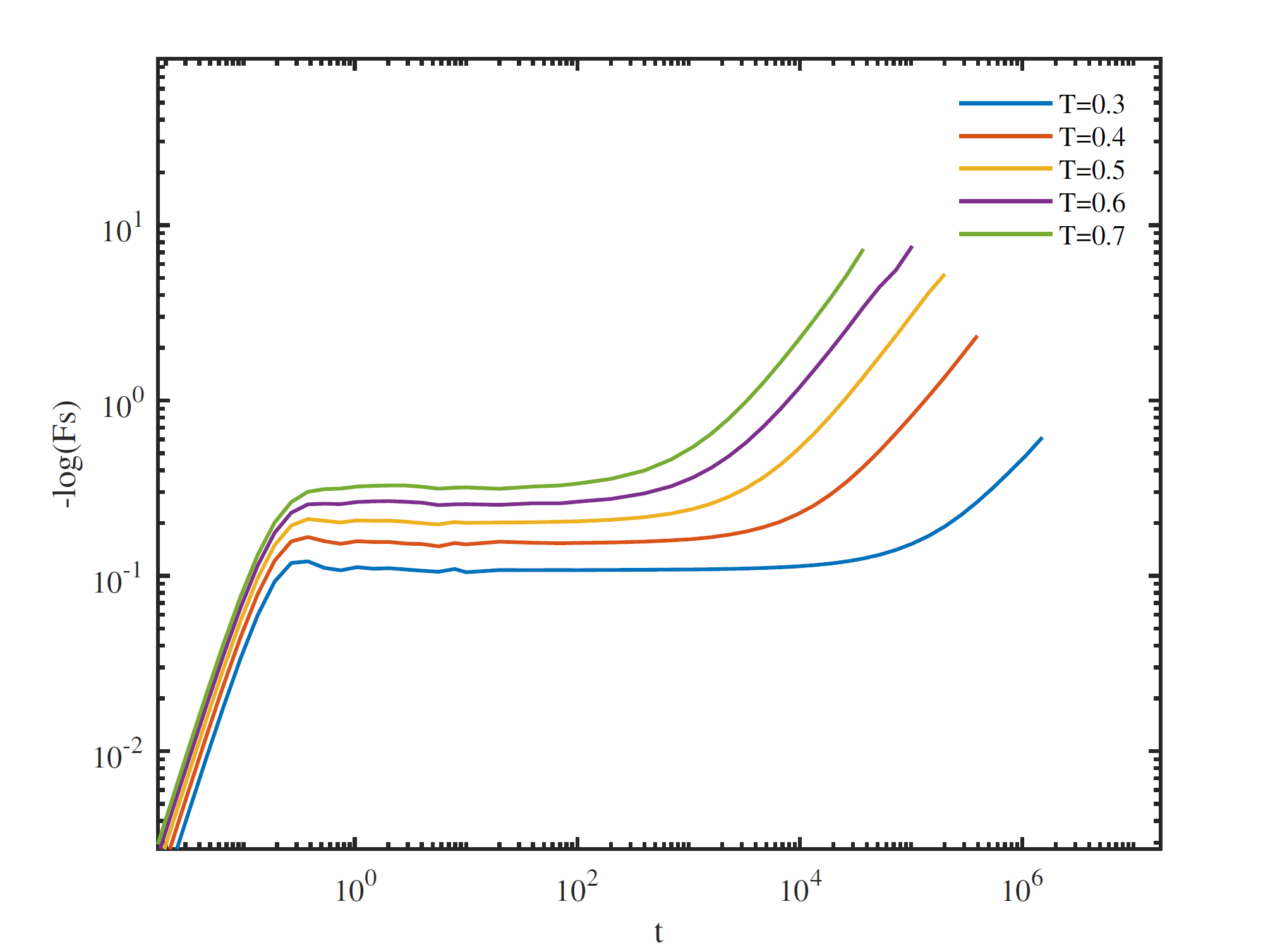}  
  \vspace{-0.07\columnwidth}
  \caption{
  Self-intermediate scattering function $F_s(q,t)$ against $t$ in a semi-log plot (a) and 
   $-\log(F_{s}(q,t))$ against t in a log-log plot (b) using the same set of data. 
  }
  \label{fig:sisf1}
\end{figure}

\begin{figure}[tb]
  \flushleft{~(a)}\\ \vspace{-0.1\columnwidth} ~~~~~~
  \includegraphics[width=\linewidth]{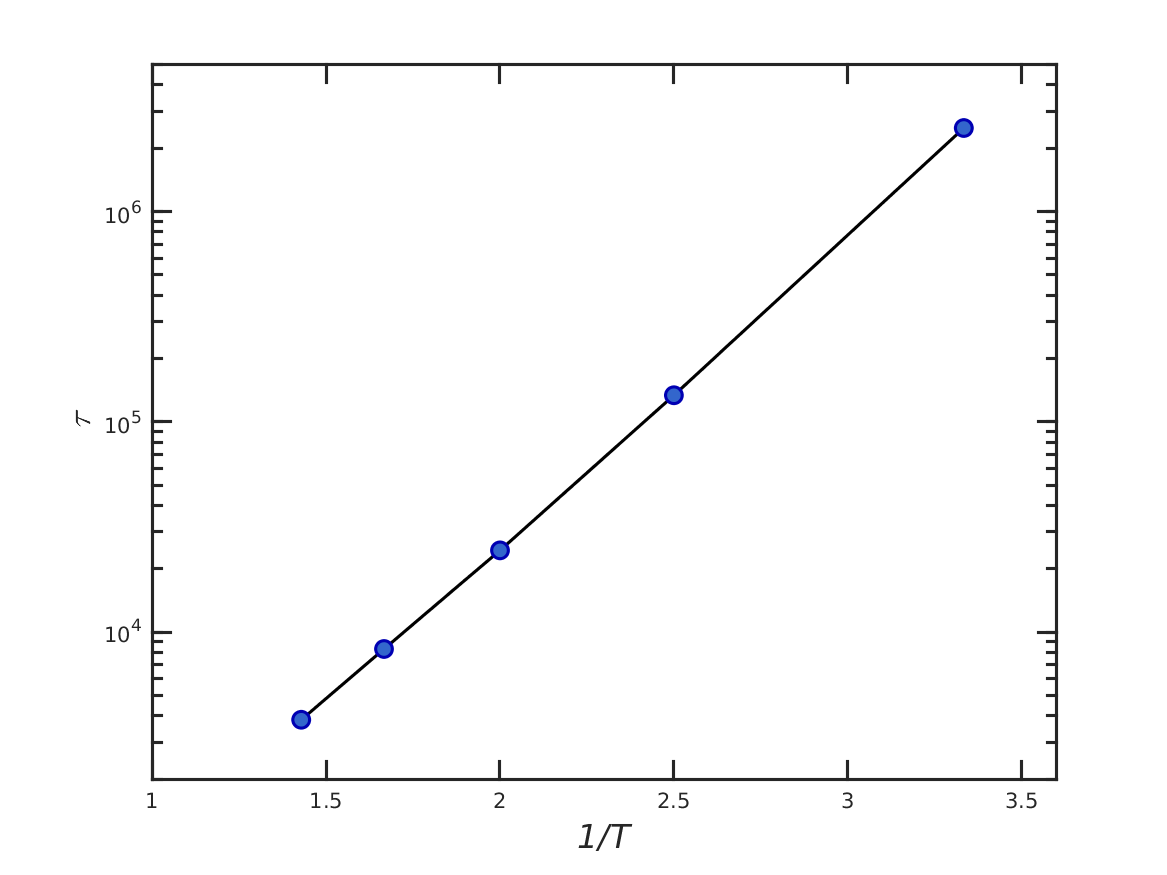}
  \vspace{-0.04\columnwidth}
  \flushleft{~(b)}\\ \vspace{-0.1\columnwidth} ~~~~~~
  \includegraphics[width=\linewidth]{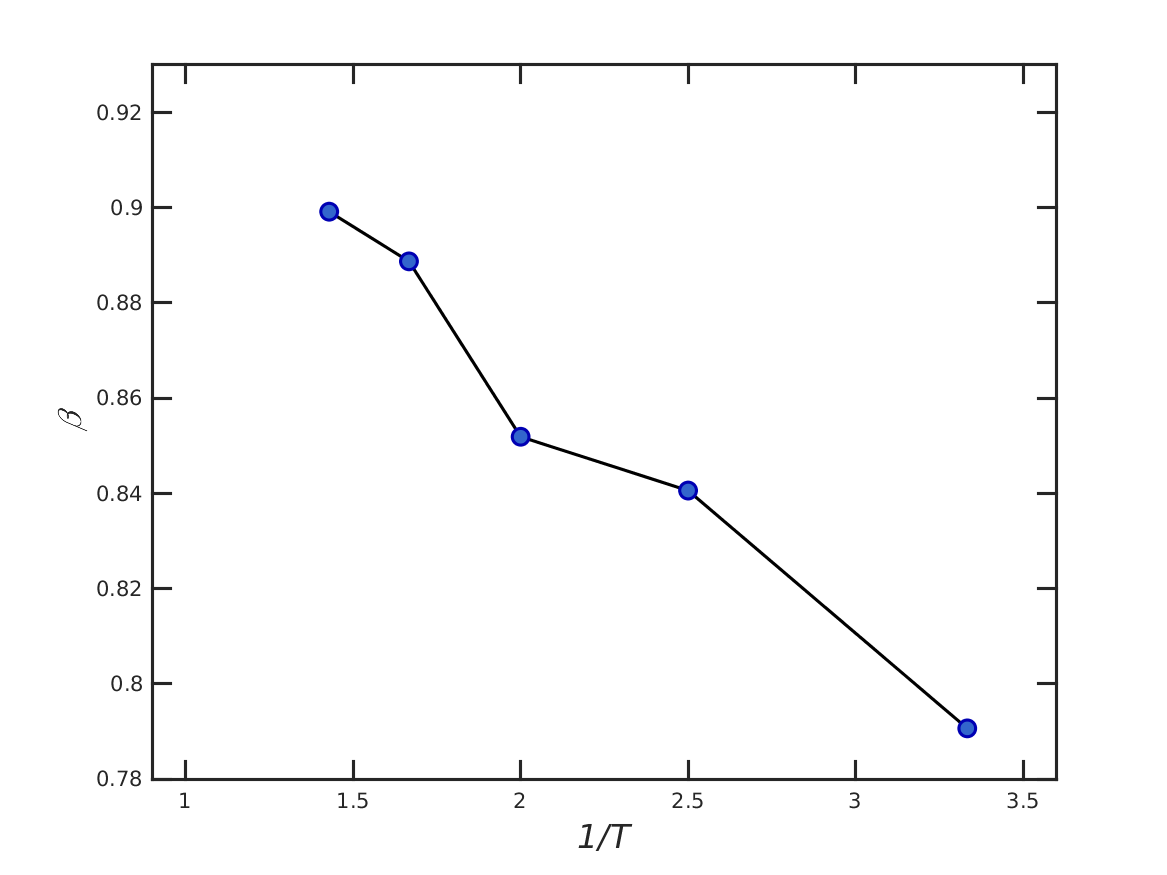}  
  \vspace{-0.07\columnwidth}
\caption{(a) Structural relaxation time $\tau$ and (b) stretching exponent $\beta$ against $1/T$ extracted from the self-intermediate scatter function \rev{for $T=0.3$ to $T=0.7$}.}
\label{fig:beta}
\end{figure}
  
\hd{Self-intermediate scattering function}
Concerning equilibrium properties, we have measured the self-intermediate scattering function (SISF), which is defined as 
\begin{equation}
F_{s}(q,t)=\left\langle e^{i\mathbf{q}\cdot\left(\mathbf{r}_{l}(t)-\mathbf{r}_{l}(0)\right)}\right\rangle,
\end{equation}
where $|\mathbf{q}|=2\pi/\lambda$ with $\lambda=a_0/\sqrt{2}$ being the nearest neighbor distance.
Results are plotted in \fig{fig:sisf1}(a) showing a two-step relaxation. 
The second terminal decay at $t\agt 10^3$ is well approximated by the Kohlraush-Williams-Watts (KWW) stretched exponential function of the form $A \exp [-(t/\tau^{\beta})]$, where $\tau$ is the structural relaxation time, $\beta (0<\beta<1)$ is the stretching exponent and decay amplitude $A$ decreases from around 0.9 to 0.7 with ascending $T$. The fit to the KWW form is demonstrated by a linear region at large $t$ in the log-log plot of $-\log(F_{s}(q,t))$ against $t$ in \fig{fig:sisf1}(b). \Fig{fig:beta}(a) shows $\tau$ against $1/T$. A linear behavior in the semi-log plot again indicates a strong glass. The stretching exponent $\beta$ shown in \fig{fig:beta}(b) is less than 1 and decreases with $1/T$. Note that the data for $T=0.3$ have been taken over a rather short duration slightly less than the relaxation time, values analyzed in this case should admit larger errors.

\begin{figure}[tb]
  \includegraphics[width=\linewidth]{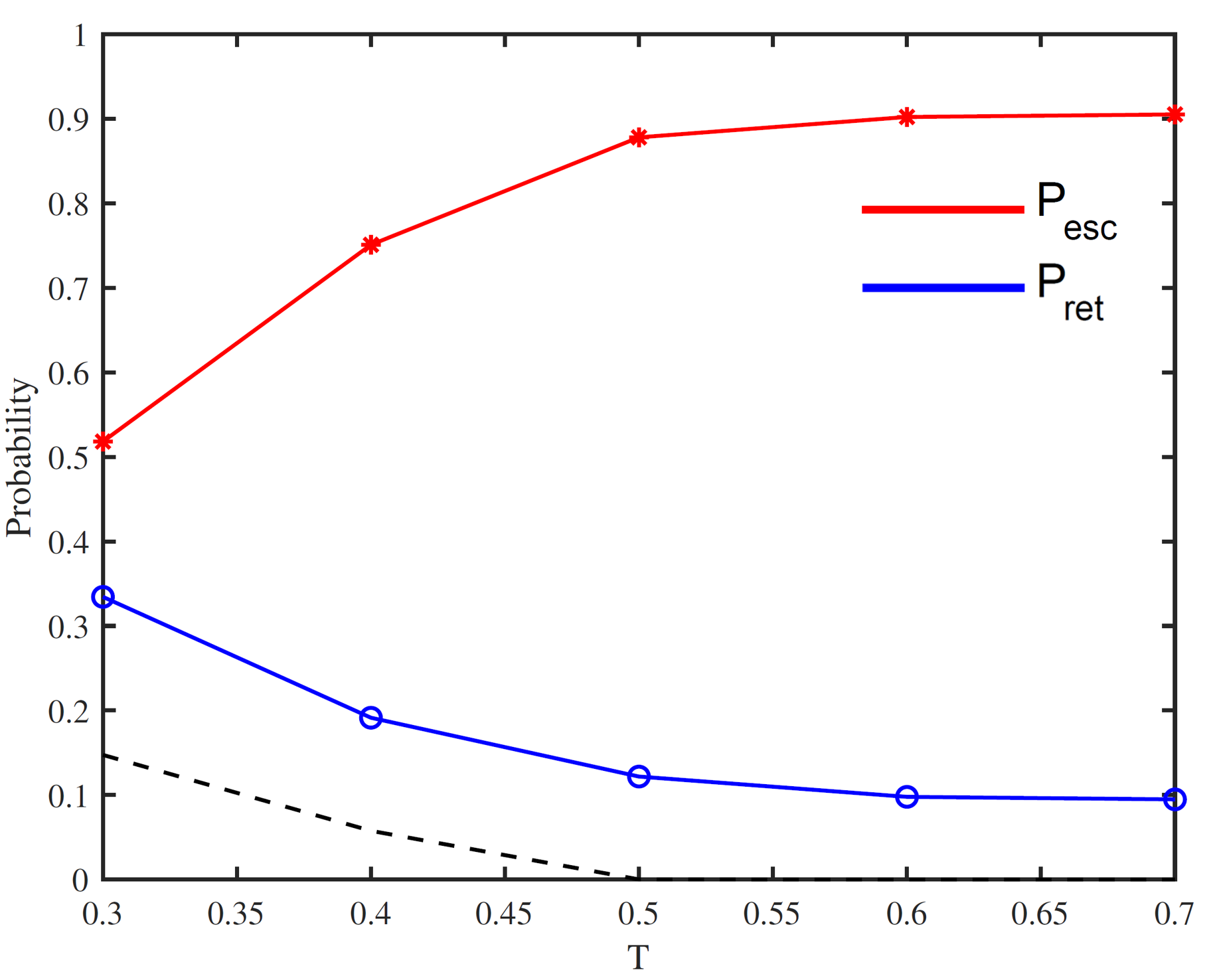}
  \caption{ Probabilities $P_{ret}$ and $P_{esc}$ for returning and escaping second hops of particles after previous hopping events against temperature $T$. The black dashed line indicates the probability that no second hop occurs during the observation period.}
  \label{fig:hopreturn}
\end{figure}

\hd{Returning and escaping hops}
As temperature decreases, dynamics slow down not only because of the reduced particle hopping rate but also because of an increased back-and-forth tendency in the hopping motions due to the rugged potential energy landscape.
We now quantify this anti-correlation in successive hops of a particle following Refs. \cite{lam2017,yip2020}. Specifically, after a particle has hopped, we measure the probability $P_{ret}$ that its next hop returns itself to the original lattice point. The probability $P_{esc}$ that it hops next instead to a new lattice position is also measured.
\Fig{fig:hopreturn} shows the results.
At a high  $T=0.7$, we find that $P_{ret} \simeq 0.095$, which is close to 0.0833 for  an uncorrelated random walk on the FCC lattice. As T decreases, $P_{ret}$ increases monotonically and reaches 0.33 for the lowest $T=0.3$ studied. This shows a strong anti-correlation in the hopping events, revealing the impact of the rugged potential energy landscape due to the random pair interactions. \rev{Analogous back-and-forth motions have also been observed in MD simulations concerning particle hops in disordered molecular systems \cite{vollmayr2004,lam2017} and rotations in orientational glassy crystals \cite{gebbia2023}.}

\hd{Computational efficiency}
\rev{The DPGC is designed to illustrate a distinct type of molecular glass model, but its computational efficiency is admittedly lower than that of standard disordered molecular models of glass or lattice models. Both the diffusion coefficient $D$ and relaxation time $\tau$ cover over roughly two orders of magnitude in the studied temperature range. The upper temperature bound is set by the FCC melting temperature $T = 0.83$, a constraint absent in disordered models, while the lowest temperature $T = 0.3$ already requires up to four months of computation with multiple CPU cores. Exploring significantly lower temperatures is currently unfeasible due to these limitations. Instead, the main advantage of the DPGC is in its simplicity and solvability.}

\rev{Despite the narrow temperature range studied, there are indications that our system at the lowest of $T=0.3$ simulated may correspond qualitatively to deeply supercooled regimes in MD simulations of disordered molecular models. The DPGC demonstrates clear plateau in MSD at $T=0.3$ spanning nearly five decades in time, as shown in Figure 2. We have also shown in Figure 8 that the return hop probability exceeds 0.3 at $T = 0.3$, which is significantly larger than the random-walk value of 1/12 for an FCC lattice, indicating a significant dynamical slowdown driven not merely by a reduced hopping rate. In addition, as forementioned, the simulations at $T = 0.3$ already required extensive computational resources. While the temperature range is narrower than in some experimental or model systems, we have pushed the simulations to a practical limit and a temperature comparable to other MD models where key glassy signatures are evidently observed.}
  
\section{Dynamical heterogeneity}
\label{sec:dyn}

\begin{figure}[tb]
  \includegraphics[width=\linewidth]{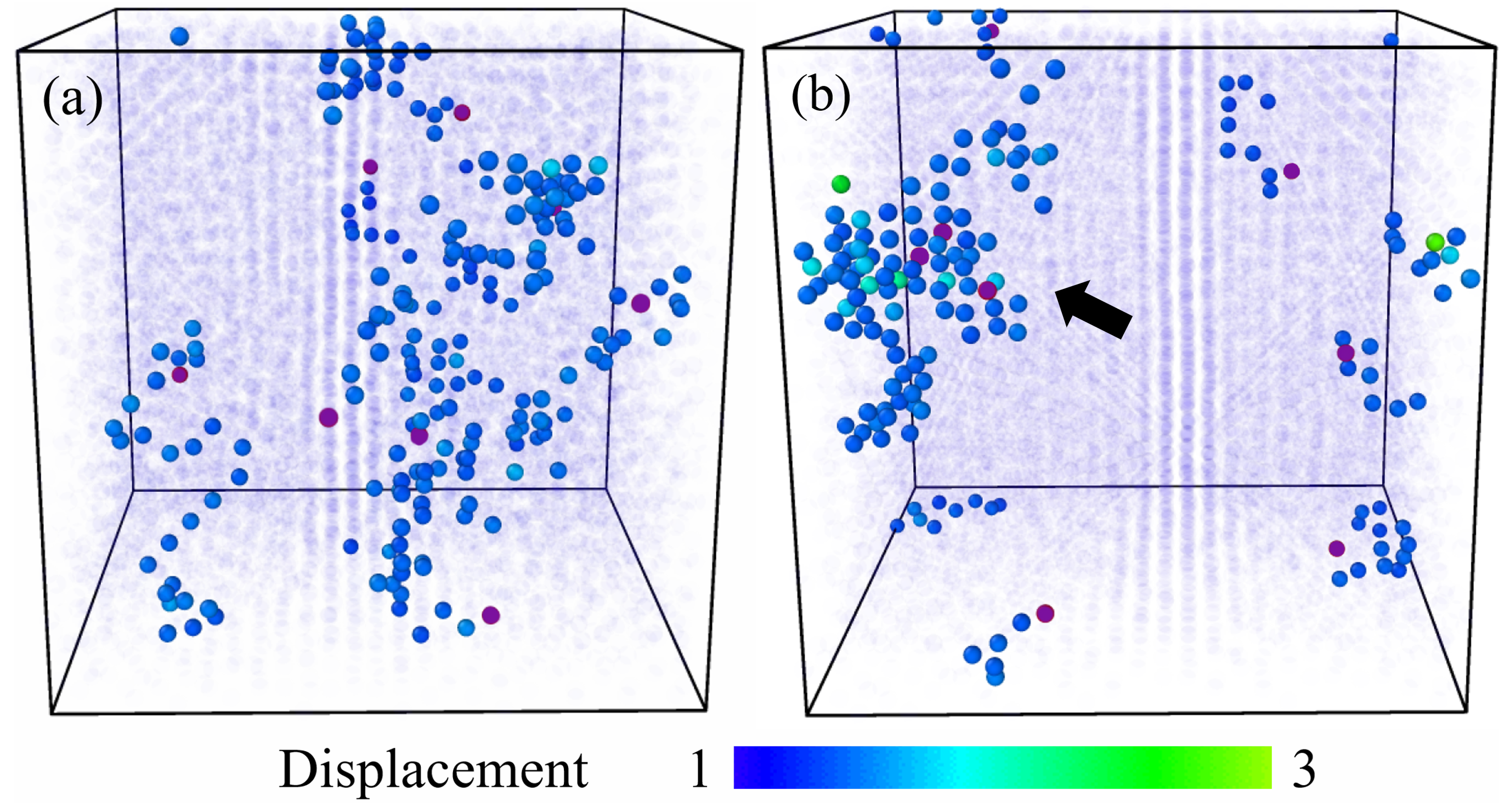}
  \caption{ Particles with large displacements from 1 (blue) to 3 (green) at $T$=0.7 (a) and 0.25 (b). Displacements are measured over a duration at which the particle MSD equals 0.1$\sigma$. Purple spheres represent vacancies. A black arrow in (b) indicates a region with a strong facilitation among vacancies.
  }
  \label{fig:trajectory}
\end{figure}

Dynamic heterogeneity is the behavior in glass formers that some regions relax much faster than the others. We now show that the DPGC exhibits dynamical heterogeneity, by real space illustrations and quantitative studies. It is easy to understand in general that dynamical heterogeneity is observed for systems following defect-induced dynamics including glassy or simple crystals with vacancies as well as partial-swap systems \cite{gopinath2022}. The idea can also apply to conventional glasses if one assumes quasivoid-induced dynamics \cite{yip2020}. 

To illustrate the heterogeneity in real space, \fig{fig:trajectory} highlights particles in the DPGC with large displacements.
As seen, they concentrate close to the vacancies. The vacancy-induced nature of particle hops immediately explains the dynamic heterogeneity. It also explains the observed stringlike geometries of the set of hopping particles \cite{glotzer1998,kawasaki2013}, revealing the paths taken by the vacancies.
Note that vacancy-induced motion realizes a defect-particle facilitation process \cite{fredrickson1984,kob1993,lam2017,zhang2017}, which also mediates facilitation among particles \cite{ganapathy2014,scalliet2022}.

At $T=0.7$ [\fig{fig:trajectory}(a)], various vacancies have rather similar mobilities.
As $T$ decreases to $0.25$ [\fig{fig:trajectory}(b)], isolated vacancies are observed to slow down more significantly, and mobility is dominated by groups of vacancies. 
Specifically, when vacancies are energetically attracted to each other, adjacent particles become loosely bonded and hop more rapidly, speeding up locally the motions of both the particles and the vacancies. This realizes a form of defect-defect facilitation mainly of an energetic origin. Particles close to groups of vacancies thus enjoy enhanced dynamics and this increases the dynamic heterogeneity at low $T$. Nevertheless, such facilitation is distinct from facilitation of a dynamic origin in the absence of defect attraction \cite{fredrickson1984,kob1993,lam2017} as has been observed in the DPLM \cite{zhang2017,lam2018tree}. Further work in the future is required to control the vacancy-vacancy attraction based on alternative particle potentials so as to implement facilitation better resembling that in conventional glasses. \rev{We should also point out that there have also been many important studies on dynamic facilitation without explicit defects \cite{garrahan2011review,chandler2011,isobe2016}, which provide possible descriptions of disordered molecular systems as well as vacancy-free orientational glassy crystals.   }

\begin{figure}[tb]
  \includegraphics[width=\linewidth]{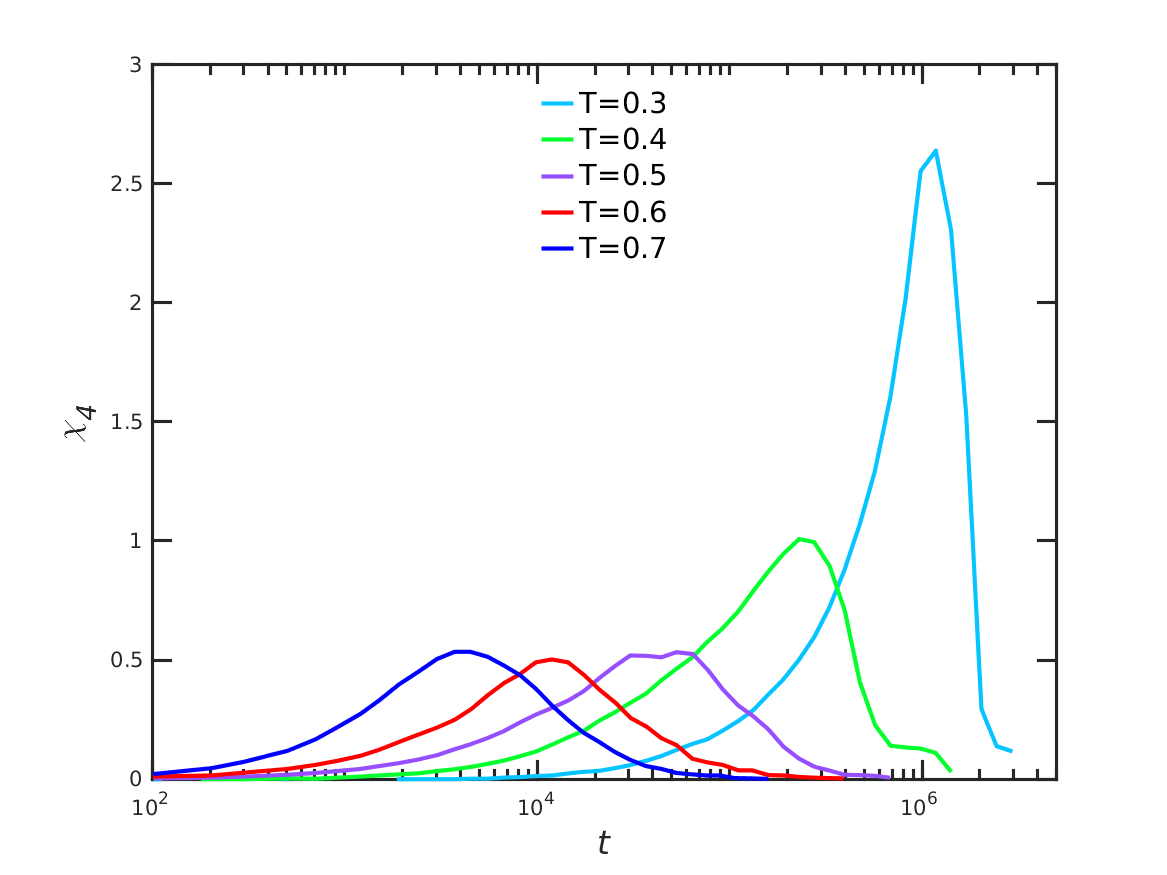}
\caption{Four-point susceptibility $\chi_4(t)$ against time $t$.}
\label{fig:chi4}
\end{figure}

To quantitatively study dynamic heterogeneity, one first defines an overlap function as
\begin{equation}
    c_l(t,0) = e^{i \mathbf{q} \cdot \left(
        \mathbf{r}_l(t) - \mathbf{r}_l(0)
    \right)}.
\end{equation}
where $|\mathbf{q}|=2\pi/\lambda$ with $\lambda=a_0/\sqrt{2}$. 
Note that the average overlap equals the SISF $F_s({q}, t)$. Each particle contributes to an overlap field defined by
\begin{equation}
    c(\mathbf{r};t,0) = \sum_l c_l(t,0) \delta\left( \mathbf{r} - \mathbf{r}_l(0) \right),
\end{equation}
where the sum is over all particles $l$. Consider its spatial correlation
\begin{equation}
    G_4(\mathbf{r},t)
    = \left\langle  c(\mathbf{r};t,0) c(\mathbf{0};t,0) \right\rangle -
      \left\langle c(\mathbf{0};t,0) \right\rangle^2
\end{equation}
where the average is over the spatial origin $\mathbf{0}$ and the starting time 0. Then, $G_4$ measures the correlation of the fluctuations in the overlap function between two points that are separated by $\mathbf{r}$. In the Fourier space, we get
\begin{eqnarray}
    S_4(\mathbf{\tilde q},t)
    &=&  \int e^{i \mathbf{\tilde q} \cdot \mathbf{r}} G_4(\mathbf{r},t) d\mathbf{r} \\
    &=&  N \left\langle \left|
            \frac{1}{N}
            \sum_l e^{i \mathbf{\tilde q} \cdot \mathbf{r}_l(0)}
            \left(
                c_l(t,0) - F_s({q},t)
            \right)
        \right|^2 \right\rangle\nonumber\\
\end{eqnarray}
We then define the susceptibility as $\chi_4(t) = \lim_{\tilde q \to 0}         S_4(\mathbf{\tilde q},t)$, which is simply the variance of the overlap function. One can interpret $\chi_4(t)$ as the typical size of correlated clusters in structural relaxation and it is thus a measure of the degree of dynamic heterogeneity.

\Fig{fig:chi4} shows $\chi_4(t)$ measured from the simulations. As is typical for structural glass, $\chi_4(t)$ has a peak for each temperature, which shifts to larger times and has a larger height when $T$ decreases. This reveals an increasing length scale of dynamic heterogeneity when the system cools down.

\begin{figure}[tb]
\includegraphics[width=\linewidth]{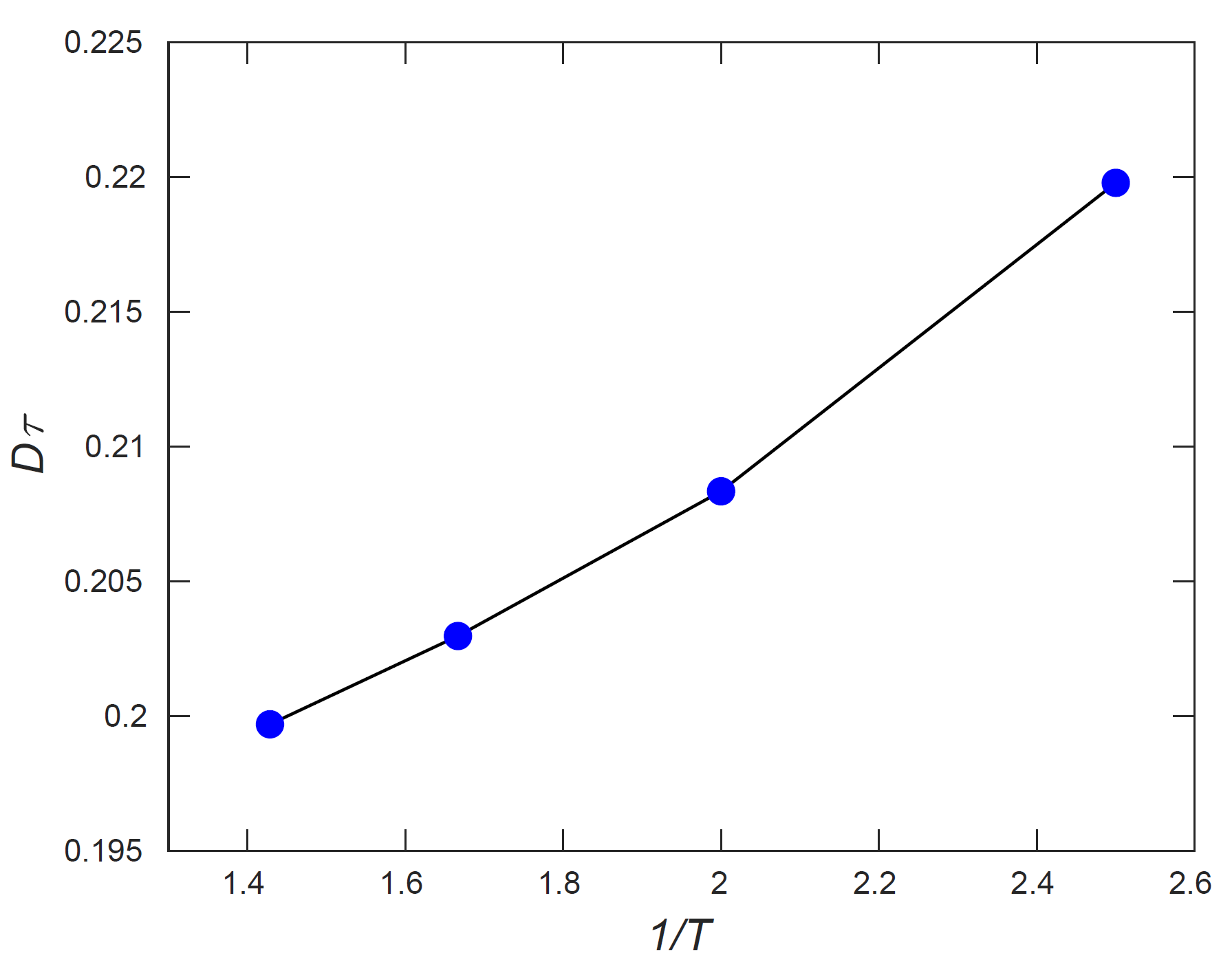}
\caption{Plot of $D\tau$ against $1/T$. A non-constant value indicates a violation of the Stokes-Einstein relation. }
\label{fig:Dtau}
\end{figure}

\Fig{fig:Dtau} plots $D \tau$ against $1/T$. We observe that $D\tau$ increases with decreasing $T$, demonstrating a violation of the Stokes-Einstein relation as expected for glasses with dynamic heterogeneity. The violation is nevertheless slight as we cannot cover a wide temperature range.

\section{Inherent structures and Equilibrium Statistics}
\label{sec:eq}

\begin{figure}[tb]
\includegraphics[width=\linewidth]{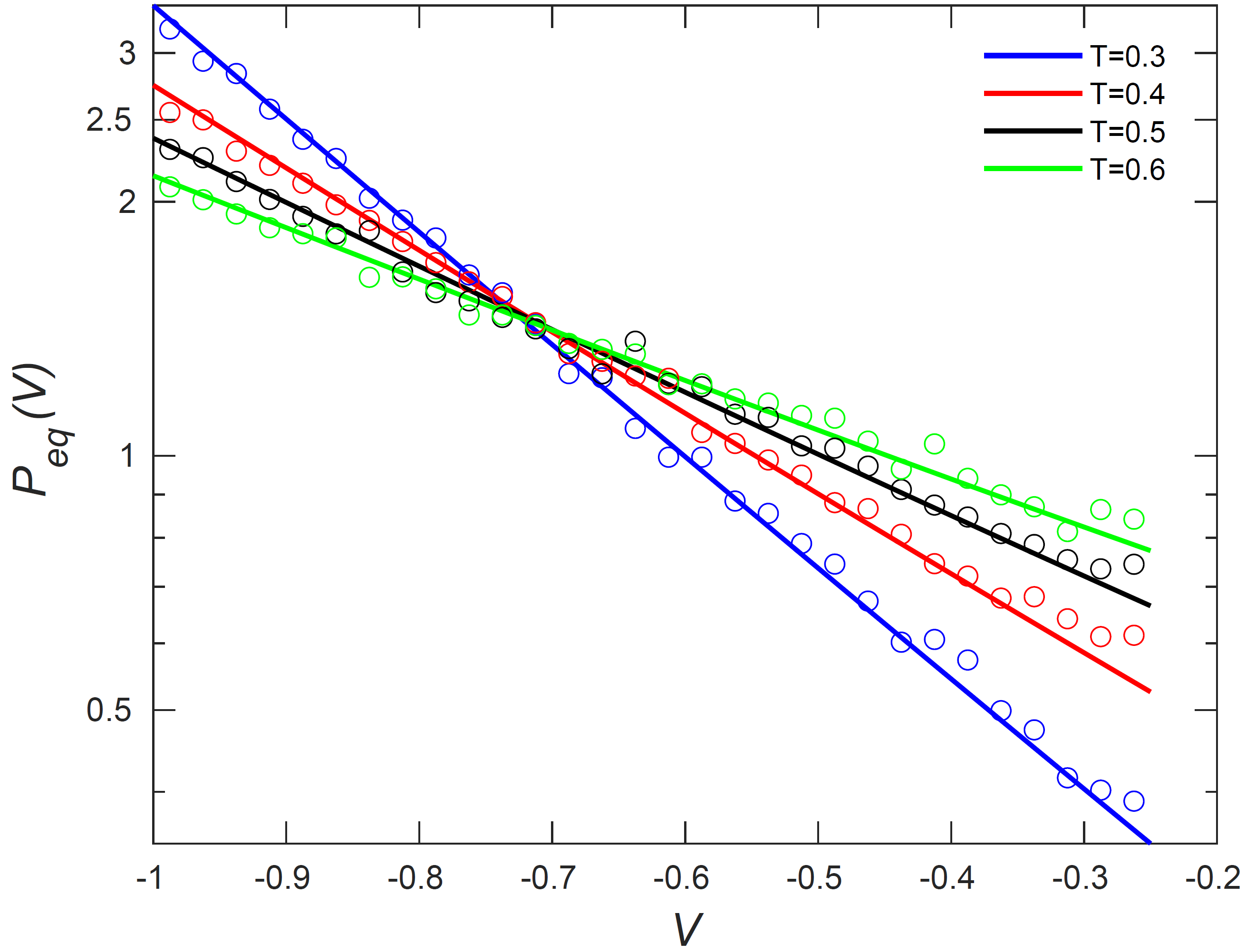}
\caption{Equilibrium distribution $P_{eq}(V)$ of pair interaction energy depth $V$ between neighbor particles from simulations (symbols) and from \eqs{peq}{Psi} (lines). }
\label{fig:peq}
\end{figure}

Glassy dynamics can be formulated in terms of transitions among inherent structures, which are meta-stable states of the system \cite{stillinger1983}.
Performing an energy minimization numerically, the DPGC arrives always at an inherent structure which is simply a FCC lattice of a certain particle and vacancy arrangement with vibrations suppressed, as shown in 
\fig{fig:lattice}. 
The number of possible 
inherent structures is $(N+N_v)!/N_v!$, noting the $N$ distinguishable particles and $N_v$ identical vacancies.

The free energy $\FIS$ of an inherent structure can be expressed as $\FIS = -k_B T \ln \PIS$, where $\PIS$ is its occurrence probability.
Analogous to exact equilibrium statistics of the DPLM \cite{zhang2017,lee2020},
we propose that $\PIS$ can be approximately factorized over the interactions and is given by  %$\prod_{kl}   P_{eq}(V_{kl})$
\begin{equation}
  \label{P}
  \PIS = \prod_{kl}   P_{eq}(V_{kl}), 
\end{equation}
where the product is over all nearest neighboring particles $k$ and $l$. Here, $P_{eq}(\Vkl)$ is the equilibrium probability of a pair interaction given by (see \app{equilstat} for derivation)
\begin{equation}
  \label{peq}
  P_{eq}(\Vkl) = \frac{1}{\N} e^{-\Vkl/k_BT}  g(\Vkl) \Psi(\Vkl) 
\end{equation}
where $\Psi(V)$ 
accounts for energetic and entropic effects of vibrations and can be approximated as
\begin{equation}
  \label{Psi}
  \Psi(\Vkl) = \left(-\Vkl- 3\bar{V}/2 \right)^{-\frac12}
\end{equation}
with $\bar{V}$ being the mean value of $\Vkl$ and
$\N$ following from the normalization $\int P_{eq}(V) dV = 1$. 
For the DPLM without vibration, $\Psi(\Vkl)\equiv 1$ and \eq{peq} is exact \cite{zhang2017, lee2020, lulli2020}.
For the DPGC, we have
measured $P_{eq}(\Vkl)$ based on the inherent structures from equilibrium simulations. Results are plotted in \fig{fig:peq} showing a good agreement with \eqs{peq}{Psi}.

\section{Discussions}
\label{sec:Conclusion}
The DPGC possesses no structural disorder. It implements directly an energetic disorder via random interactions which usually results instead from a structure disorder. Its ability to exhibit glassy properties suggests that energetic disorder may play a more direct and essential role than structural disorder in glassy dynamics. Studying structurally ordered glasses, like the DPGC, can be a much simpler  step to understand glass. Note that spin glass \cite{mezard1990book,zhou2017} also directly assumes an energetic disorder, which however is quenched in the real space. Instead, the disorder in the DPGC and the DPLM is quenched only in the configuration space \cite{zhang2017}, making it appropriate for structural glass.

Being a straightforward molecular generalization of the DPLM, properties reported here for the DPGC in general are inherited from and are closely analogous to those of the DPLM, except those related to lattice vibrations.
The realization of the DPGC strongly supports the physical relevance of the DPLM.
Conversely, we expect that glassy phenomena already demonstrated by the DPLM \cite{zhang2017,lee2021,lulli2020,lee2020,gopinath2022,  lulli2021,gao2022,gao2023,ong2024,zhai2024} likely apply also to the DPGC. These features of the DPLM support the idea that the DPGC describes a typical glass rather than a new type of glass.

Our system exhibits three phases. As $T$ increases from below $T_g$, it crosses over from the glass phase, in which particles hardly hop within practical observation times, to the dynamic phase, in which the system readily relaxes via particle rearrangements among the lattice positions. The static structures of both phases follow the same FCC lattice. At higher $T$, it melts into the liquid phase. The glass and dynamic phases of the DPGC separated by a glass transition generalize the glass and supercooled-liquid phases of conventional glass formers. The DPGC is also distinct from conventional non-glassy crystals with non-random pair interactions. While some dynamic heterogeneity can occur in such simple crystalline solids with vacancies, such  models cannot exhibit glassy features such as energy hysteresis with a kink during a temperature cycle or a tunable fragility as demonstrated by the DPGC.

The complete set of inherent structures \cite{stillinger1983} is known with the occurrence probabilities $\PIS$ and free energy $\FIS$ given in \eqr{P}{Psi}. This is, in our knowledge, unique in all molecular glassy systems, including orientational glassy crystals \cite{hochli1990review}.
Furthermore, the known $\FIS$ also directly implies a full knowledge of the potential energy landscape (PEL) \cite{stillinger1995} expressible as a function of inherent structures.
It is a rugged PEL due to random pair interactions, in sharp contrast to simple crystals and assumptions in early defect theories \cite{glarum1960}.

The elementary motions in the relaxation of the DPGC are vacancy-induced particle hops. This is analogous to deeply supercooled liquids in which quasivoids-induced  particle hops may dominate \cite{yip2020}. With the full knowledge of the PEL, a transition state theory of the dynamics \cite{vogel2004} can be straightforwardly formulated. For example, with $N_v$ monovacancies, an inherent structure is directly connected by possible transitions to $12 N_v$ others, noting that there are 12 possible hopping directions of each vacancy.
In the context of the DPLM, such an analysis implies that the energetically favorable domain of the PEL takes a random-tree geometry in the configuration space, leading to emergent kinetic constraints and facilitation \cite{lam2018tree,deng2019}. Implications to the DPGC will be studied in the future.

An accurate experimental realization of the DPGC may be challenging. One complication concerns the random particle-dependent interactions in which interaction depth between particles $k$ and $l$ is uncorrelated to that between particles $k$ and $l'$ for $l \neq l'$. 
However, allowing for correlations among the interactions, a 2D lattice of polydispersed colloidal particle system has shown signs of glassy properties \cite{chan2022}. 
Alternatively, limiting to few particle types, our system is analogous to high-entropy alloys \cite{yeh2013}. The DPGC can serve as an idealized model for studying various types of glass.

In conclusion, we have developed a distinguishable-particle glassy crystal model which  shows typical glassy behaviors. The inherent structures are known and numerable with their approximate equilibrium probabilities available analytically.
This makes it, in our opinion, the simplest molecular model of glass. It also demonstrates that a structural disorder is not essential in the presence of an energetic disorder to exhibit glassy properties.

\section{Acknowledgments}

We are grateful to Giorgio Parisi and Paolo Balden for helpful discussions.
This work was supported by General Research Fund of Hong Kong (Grant 15303220), National
Natural Science Foundation of China (Grant No. 1217040938, 12174079 and 11974297).

\section{Author Declarations}

\subsection{Conflict of Interest}
The authors have no conflicts to disclose.

\section{Data Availablity Statement}

The molecular dynamics scripts and the data analysis scripts of this article are available at https://github.com/leolamsi/gc.

\appendix
\section{Melting}
\label{melting}
\begin{figure}[tb]
\includegraphics[width=\linewidth]{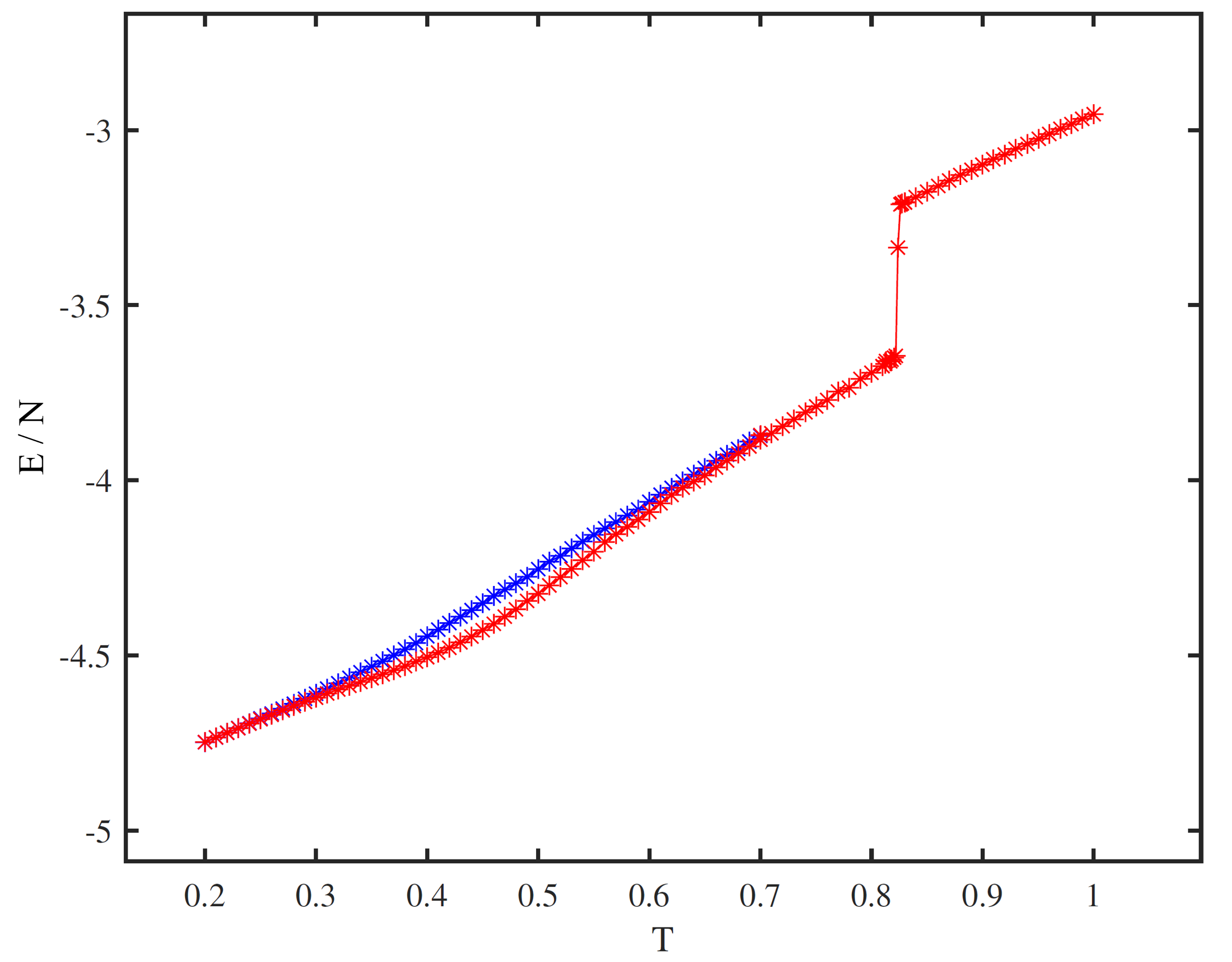}
\caption{
Potential energy per particle $E/N$ against temperature $T$ during cooling (blue) and heating (red) with a rate $\nu= 10^{-6}$. The abrupt jump at $T\simeq 0.83$ signifies melting of the FCC lattice.
}
\label{fig:E_T_melt}
\end{figure}

To determine the temperature range for the energy hysteresis shown in \Fig{fig:cool}, we have heated the glassy crystal to a higher temperature as shown in  \fig{fig:E_T_melt}. Akin to the energy hysteresis, we have performed a heating and cooling cycle at a cooling/heating rate $\nu=10^{-6}$. At $T \simeq 0.83$, an abrupt  jump of the energy indicates the melting of the FCC lattice, as the system becomes structurally disordered after the transition. All our main simulations are performed well below 0.83 to make sure the FCC lattice is stable.

\section{Initialization via particle/vacancy swap}
\label{fastequil}
System equilibration using standard MD steps can take a very long runtime at low $T$. This can be dramatically improved using particle swaps \cite{ninarello2017} in addition to the MD steps
at not too low $T$. However, at the lowest $T=0.3$ studied, we have found surprisingly that particle swaps fail to equilibrate our system completely. A close examination reveals that even though the distribution of the pair interaction depth $V$ has already converged to $P_{eq}(V)$, the density of di-vacancies is not at equilibrium after particle swaps. This is because the method does not relax the positions of the vacancies. Due to the considerable attraction of the vacancies, the density of di-vacancies should increase as $T$ decreases but this is not achieved using swaps. For conventional glass models, quasivoids, the counterpart of vacancies, appear unattractive to each other and the problem thus does not apply.

To fully equilibrate the DPGC at $T= 0.3$, we have developed a particle/vacancy swap algorithm in addition to the MD steps. For convenience of implementation in LAMMPS, we represent vacancies using $ghost$ particles. 
Specifically, we fill up each vacancy position with a ghost particle, which has a light mass $m_{ghost}=0.01$ and a weak interaction depth  $V_{ghost}=-0.01$ with all other particles. Then, they introduce negligible perturbations to the original crystal. 
A small timestep of $0.0001$ is used in the MD steps to avoid the ghost particles being bounced off the system.

The full algorithm for $T=0.3$ is as follows. We randomly arrange all real and ghost particles in the FCC lattice. We then perform non-local pairwise particle swaps, irrespective of whether they are real or ghost. This is done as usual by choosing any two particles and swapping their positions with a probability $\exp(-\Delta{E} /k_{B}T)$, where $\Delta{E}$ is the energy change after the swap. The swapping process is performed periodically in between normal MD steps. When equilibrium is attained, as indicated for example by the stabilization of the potential energy, the ghost particles are removed. The system is then further relaxed using conventional swap and MD steps as a safety precaution.
We have found that these procedures successfully equilibrate the DPGC at $T=0.3$.

\section{DPGC as a fragile glass}
\label{fragile}

We have demonstrated that the glassy crystal discussed in the main text exhibits strong glass relaxation behaviors as a uniform interaction energy distribution $g(V)$ is applied within the range $V \in [V_0,V_1] \equiv [-1, -0.25]$. Here, we show that the model can also simulate a more fragile glass. Following Ref. \cite{lee2020} in fine-tuning the fragility of the DPLM, we apply a uniform-plus-delta bi-component form 
\begin{equation}
    g(V) = \frac{G_0}{\Delta V} + (1 - G_0)\delta(V - V_1),
\end{equation}
where $\Delta V = V_1 - V_0 = 0.75$ and $ \delta $ denotes the Dirac delta function.  For $G_0=1$, $g(V)$ reduces back to the uniform distribution studied above for strong glass.

\begin{figure}[tb]
  \includegraphics[width=0.9\linewidth]{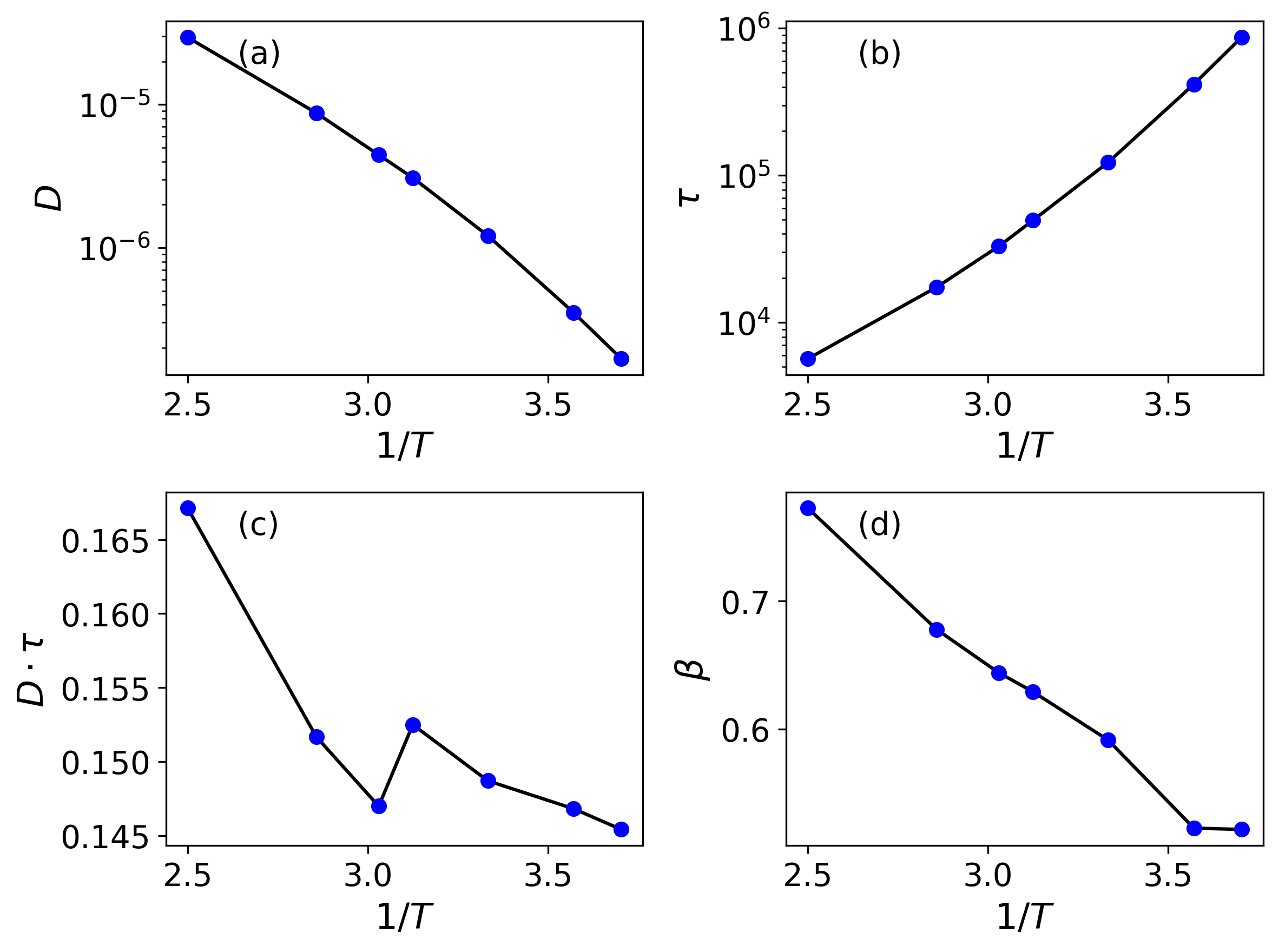}
  \caption{Results for the system with $G_0 = 0.3$ for $T=0.27$ to $T=0.4$. (a) The super-Arrhenius relationship demonstrates fragile-type relaxation behavior. (b) Structural relaxation time $\tau$ as a function of $1/T$. (c) A non-constant value indicates a violation of the Stokes-Einstein relation. (d) Stretching exponent $\beta$ as a function of $1/T$.}
  \label{G03}
\end{figure}

We now consider the case of $G_0 = 0.3$, which has led to a more fragile glass for the DPLM \cite{lee2020}.
\Fig{G03} shows the results for the DPGC with $G_0 = 0.3$, analyzed using the same methods as explained in the main text. The melting point in this case is roughly 0.45 so that  only results for $T \le 0.4$ are reported.
As seen, the diffusion coefficient $D$ and the relaxation time $\tau$ follow super-Arrhenius temperature dependence, indicating a fragile glass. The results also indicate a small Stokes-Einstein violation and a stretching exponent $\beta$ decreasing as temperature decreases. In particular, $\beta$ reaches a small value of 0.52, below that of the strong glass, similar to findings in the DPLM \cite{lee2020}.

\section{Equilibrium statistics}
\label{equilstat}

 \begin{figure}[tb]
  \includegraphics[width=\linewidth]{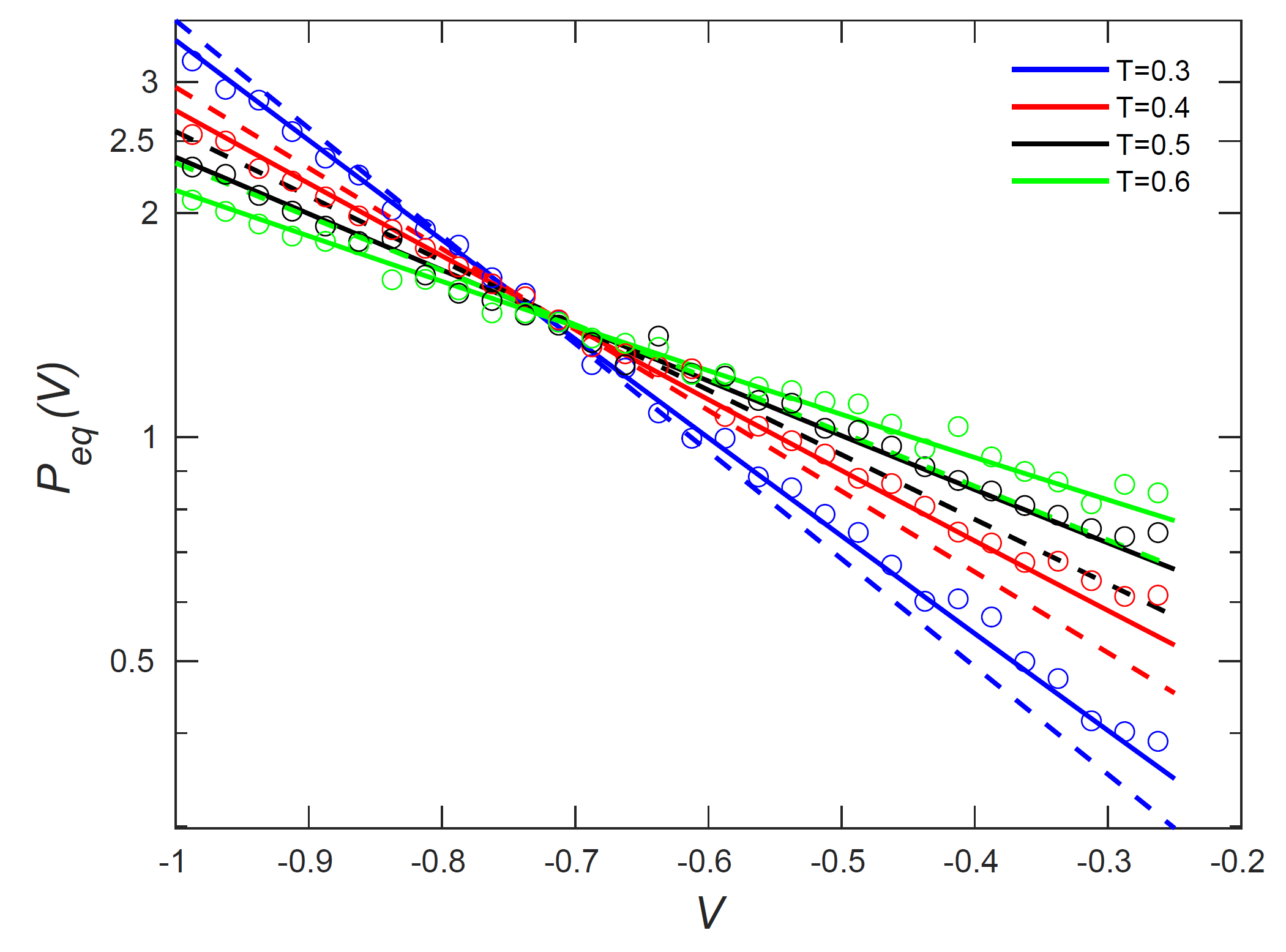}
  \caption{
Equilibrium distribution $P_{eq}(V)$ of pair interaction depth $V$ from simulations (symbols) using data from \fig{fig:peq} compared \eq{peq0} without considering vibration (dashed lines) and \eqs{peq}{Psi} with vibrations (solid lines).
}
  \label{fig:peq0}
\end{figure}
Approximate equilibrium statistics of the DPGC can be calculated by generalizing an exact method used for the DPLM.
Specifically, we study the equilibrium distribution $P_{eq}(V)$ of the energy depth $V$ of the pair interactions realized between neighboring particles. Measured values of $P_{eq}(V)$ are shown in \Fig{fig:peq} and reproduced in \Fig{fig:peq0}.

A rough estimate $P_{eq}^0(V)$ of $P_{eq}(V)$, assuming naively a simple Boltzmann weight $e^{-V/k_BT}$ following DPLM results, is  given by
\begin{equation}
  \label{peq0}
  P_{eq}^0(V) \propto e^{-V/k_BT}  g(V) .
\end{equation}
\Eq{peq0} has been proven exact for the DPLM when $V$ is taken as the pair interaction.
\Fig{fig:peq0} compares $P_{eq}^0(V)$ with the numerically measured $P_{eq}(V)$ and shows a fair agreement.

We note that the DPGC differs from the DPLM mainly by having particle vibrations.
Generalizing \eq{peq0} to  account for energetic and entropic effects of vibrations approximately, we replace the Boltzmann factor in \eq{peq0} by the partition function $Z_1(V)$ of an interaction of depth $V$ as follows 
\begin{equation}
  \label{peq1}
  P_{eq}(V) \propto Z_1(V)  g(V),
\end{equation}
where
\begin{equation}
  \label{Z1}
  Z_1(V) = \int_0^{\infty} \exp\squarebk{ -E_V(r)} dr
\end{equation}
with $E_V(r)$ being the average system energy when an interaction of depth $V$ is at bond length $r$. 
More precisely, we focus on the interaction between neighboring particles $k$ and $l$ of depth $V$ separated by a distance $r$. Assume for simplicity that only  particles $k$ and $l$ vibrate in the breathing, i.e. symmetric, mode and all other particles are stationary at the lattice positions.
The separation can be expressed in terms of particle displacement $s$ as $r = r_0+2s$
with $r_0=2^{1/6}\sigma$.
Assume that all other interactions have a uniform depth $\overline{V}$ given by its approximate average value 
\begin{equation}
  \overline{V} = \int V P_{eq}^0(V) dV.
\end{equation}
An analysis of the 23 interactions among particles $k$ and $l$ and nearest neighbors in the FCC lattice indicates that only 11 interactions depend on $s$ up to the linear order and we get     
\begin{eqnarray}
  E_V(s) &=& \Phi(V, 2s) + 2\Phi(\overline{V}, -s) 
  + 4\Phi(\overline{V}, s/\sqrt{2}) \nonumber\\
  && + 4 \Phi(\overline{V}, -s/\sqrt{2}),
\end{eqnarray}
where the Lennard-Jones potential is parametrized as 
\begin{equation}
\Phi(V,s)=-4V\left[\left(\frac{\sigma}{r_0+2s}\right)^{12}-\left(\frac{\sigma}{r_0+2s}\right)^{6}\right].
\end{equation}
For further simplification, we use a harmonic approximation for the LJ potential, i.e.
\begin{equation}
  \Phi(V,s)= -\frac12 KV \cdot (2s)^2 + V, 
\end{equation}
where $K$ is an effective elastic constant. 
\Eq{Z1} then involves a simple Gaussian integral which can be evaluated to get
\begin{equation}
  Z_1(V) = \exp\roundbk{-\frac{V+10\overline{V}}{k_BT}}\sqrt{-\frac{4\pi k_BT}{2 KV+ 3 K\overline{V}}}.
\end{equation}
Using also \eq{peq1}, we have
\begin{equation}
  P_{eq}(V) = \frac{1}{\N} e^{-\Vkl/k_BT}  g(\Vkl) \left(-V- \frac32 \overline{V}\right)^{-\frac12},
  \label{peq2}
\end{equation}
where factors independent of $V$, such as $K^{-1/2}$, have been rescaled away when defining the normalization constant $\N$.
\Eq{peq2}, equivalent to \eqs{peq}{Psi}, shows much better agreement with simulations than \eq{peq0} as shown in \fig{fig:peq0}.

The success of \eq{peq2} in describing our simulations show that interactions in a system is approximately independent of each other, which is an exact result for the DPLM. 
As a consistency check, at a very low $T$, 
$P_{eq}(V)$ is non-negligible only for $V\simeq \overline{V}$ so that the last factor in \eq{peq2} approaches a constant. Then, $P_{eq}(V)$ converges to $P^0_{eq}(V)$ as vibrations diminish.

\bibliography{glass_short}

\end{document}